# Reinforcement Learning for Personalized Drug Discovery and Design for Complex Diseases: A Systems Pharmacology Perspective

Ryan K. Tan[1], Yang Liu1, Lei Xie[1,2,3,*]

[1]Department of Computer Science, Hunter College, The City University of New York

[2]Ph.D. Program in Computer Science, Biology & Biochemistry, The Graduate Center, The City University of New York

[3]Helen and Robert Appel Alzheimer's Disease Research Institute, Feil Family Brain & Mind Research Institute, Weill Cornell Medicine, Cornell University

*Correspondence should be addressed

## Abstract

Many multi-genic systemic diseases such as neurological disorders, inflammatory diseases, and the majority of cancers do not have effective treatments yet. Reinforcement learning powered systems pharmacology is a potentially effective approach to design personalized therapies for untreatable complex diseases. In this survey, state-of-the-art reinforcement learning methods and their latest applications to drug design are reviewed. The challenges on harnessing reinforcement learning for systems pharmacology and personalized medicine are discussed. Potential solutions to overcome the challenges are proposed. In spite of successful application of advanced reinforcement learning techniques to target-based drug discovery, new reinforcement learning strategies are needed to address systems pharmacology-oriented personalized *de novo* drug design.

## 1. Introduction

Drug discovery and development is a costly and heavily time-consuming process. Despite massive investment of time and money, it is infeasible to explore the whole chemical space of drug-like compounds, which is composed of around $10^{33}$ small molecules [1], by using conventional technologies. Owing to the improvement of computer power, tremendous progress in deep learning (DL) and emergence of quantum computing, computer-aided drug design has the potential to dramatically speed up the drug discovery process. With a reasonably reliable and accurate computational model, it is possible to synthesize and test a small number of compounds precisely interacting with expected drug target(s) and achieve desirable clinical outcomes. Recently, researchers have introduced various methods to generate novel molecules or optimize existing molecules towards designed properties. The mostly used methods include generative adversarial neural-networks (GAN) [2], variational autoencoder (VAE) [3], normalizing flow [4, 5], and reinforcement learning (RL) [6]. The molecules generated by GAN, VAE, and normalizing flow are biased to specific data distributions. They generally lack the ability to explore the unknown space that has a distribution shift or directly optimize molecules toward specific targets. RL, on the other hand, is able to learn or tune a generative model specifically toward the properties of interest and enable the model to generate molecules that have different distribution from the training data. However, RL is usually less efficient compared with other methods. Training a model with RL from scratch will either cost a long time or lead to a model that is hard to converge. Thus, recent studies tend to combine pre-training or adversarial training with RL to take the advantage of the exploitation ability of transfer or adversarial learning [7–9] and the exploration power of RL.

Existing efforts in applying RL to drug discovery mainly follow conventional one-drug-one-gene target-based paradigm. Although target-based drug discovery is mostly successful in tackling mono-genic diseases whose etiologies are driven by a single gene, it suffers from high failure rate especially for multi-genic, multi-factorial, heterogeneous diseases. Moreover, a drug rarely interacts only with its primary target in human bodies. Off-target effects are common, and may contribute to therapeutic effects or side effects [10]. Therefore, systems pharmacology that targets a gene-gene interaction network instead of a single gene and is tailored to individual patients has emerged as a new drug discovery paradigm for complex diseases. However, unlike target-based compound screening that can be easily measured by drug-target binding affinities, advances of systems pharmacology are hindered by the lack of effective read-outs for high-throughput compound screening. Powered by the development of many high-throughput cell-based phenotypic detection methods, phenotype-based drug discovery starts to gain an increasing attention in recent years due to its ability to identify drug lead compounds in a physiologically relevant condition [11]. Phenotype-based drug discovery is a target agnostic and empirical approach to exploit new drugs with no prior knowledge about the drug target or mechanism of action in a disease [12]. The use of molecular signatures as phenotype read-outs makes it possible to not only establish robust drug-disease associations but also deconvolute drug targets from the unbiased phenotype screening. Additionally, phenotype-based drug discovery has the power to exploit drugs for rare or poorly understood diseases such as neurological disorders and many types of cancers. Recently, several computational methods have been developed for high-throughput phenotype compound screening using chemical-induced gene expressions [13] and images [14] as read-outs. Tremendous progress in protein structure predictions [15, 16] and development of new methods for exploring dark chemical genomics space [17] significantly enhance our ability to deconvolute genome-wide target profiles for dark proteins that are not readily accessible by experimental methods. Another fundamental challenge in systems pharmacology-oriented precision drug design is to transfer compound activity in cell-line and animal models into therapeutic efficacy in an individual patient. Although human tissue-based organoid and *ex vivo* models have been developed for anti-cancer drug testing, they are expensive and often infeasible and even unethical in many disease areas. Thus, computational approaches such as transfer learning which can predict new target variables (e.g., tumor growth) of unseen samples (e.g., patients) from a model trained by an existing data set with different distributions (e.g., cell line screens) and targets (e.g., IC50 of cell viability) can be an indispensable tool to fill in the knowledge gap between *in vitro* bioassays and *in vivo* clinical endpoints of drug candidates. These computational tools pave the way for systems pharmacology-oriented high-throughput compound screening for personalized drug discovery.

This review will be organized as follows. We will first give a brief overview of RL, including definitions of some key concepts, problem setting and formulation of leading methods. Then we will survey the recent developments of applying deep RL to drug discovery. Finally, we will highlight the challenges and opportunities of RL in systems pharmacology and personalized medicine.

## 2. Overview of reinforcement learning

In reinforcement learning, there are usually two main characters -- an agent and an environment. The agent is the key component of RL that makes sequential decisions, and the environment is the world that the agent lives in. A typical RL problem can be considered as training an agent to interact with an environment that follows a Markov Decision Process (MDP) [18].

In each interaction (with the environment), the agent receives the information of the current state $s_t \in \mathcal{S}$ and performs an action $a_t \in \mathcal{A}$ accordingly, where $\mathcal{S}$ and $\mathcal{A}$ are state and action spaces.[1] After performing an action $a_t$, the agent will transition to a new state $s_{t+1}$ and receive a reward $r_t$. These are characterized by underlying state transition dynamics $P: \mathcal{S} \times \mathcal{A} \to \Delta(\mathcal{S})$ and the reward function $r: \mathcal{S} \times \mathcal{A} \to \mathbb{R}$, i.e., $P(s_{t+1}|s_t, a_t)$ and $r(s_t, a_t)$ are the probability and reward of taking action $a_t$ in state $s_t$ and then transitioning into state $s_{t+1}$. This process repeats indefinitely or until a predefined termination condition is met. The sequence of states and actions followed in this process constitutes a so-called trajectory $\tau$, e.g., at time $t$, $\tau_t = \{s_1, a_1, s_2, a_2, ..., s_t, a_t\}$. Moreover, with a discount factor $\gamma \in (0, 1]$, we can define the discounted cumulative reward under a trajectory $\tau$ as $R(\tau) = \sum_{t=1}^{\infty} \gamma^{t-1} r(s_t, a_t)$. An MDP $\mathcal{M}$ can be represented as a tuple of all components mentioned above along with an initial state distribution $\mu$, i.e., $\mathcal{M} = \{\mu, \mathcal{S}, \mathcal{A}, P, r, \gamma\}$.

In the typical setting of MDP, agent behaves by following a (stationary) policy $\pi$, which specifies a decision-making strategy in which the agent chooses an action $a_t$ adaptively only based on its current state $s_t$. Precisely, a stochastic policy is specified as $\pi: \mathcal{S} \to \Delta(\mathcal{A})$ while a deterministic policy is of the form $\pi: \mathcal{S} \to \mathcal{A}$.

Given an MDP $\mathcal{M}$ and a policy $\pi$, we can define some functions that measures the quality of being in a state $s$ or taking an action $a$ upon $s$ in the long run. Specifically, we can define the value function $V_\mathcal{M}^\pi : \mathcal{S} \to \mathbb{R}$ that gives discounted sum of future rewards at a state $s$ following an arbitrary policy $\pi$:

$$V_\mathcal{M}^\pi(s) = \mathbb{E}_{\tau \sim T_\pi(\tau)}\left[\sum_{t=1}^{\infty} \gamma^{t-1} r(s_t, a_t) \mid \pi, s_1 = s\right] \tag{1}$$
where $T_\pi(\tau) = \prod_{t=1}^{\infty} \pi(a_t|s_t) P(s_{t+1}|s_t, a_t)$

---

[1] We use $t$ to track the time-step, e.g., $s_t$ is the state at time $t$.

Similarly, the action-value function (or Q-function) $Q_\mathcal{M}^\pi : \mathcal{S} \times \mathcal{A} \to \mathbb{R}$ can be defined as:

$$Q_\mathcal{M}^\pi(s, a) = \mathbb{E}_{\tau \sim T_\pi(\tau)}\left[\sum_{t=1}^{\infty} \gamma^{t-1} r(s_t, a_t) \mid \pi, s_1 = s, a_1 = a\right] \tag{2}$$

With $V_\mathcal{M}^\pi(s)$ and $Q_\mathcal{M}^\pi(s, a)$, we can also define another value function, the advantage function $A_\mathcal{M}^\pi(s, a) = Q_\mathcal{M}^\pi(s, a) - V_\mathcal{M}^\pi(s)$, which measures the relative reward that the agent could obtain by taking a particular step $a$ upon $s$ (compared to an average action).

The objective of RL is to learn a policy $\pi$ that optimizes the expectation of accumulated reward under a particular initial state distribution $\mu$:

$$\max_{\pi \in \Pi} J_\mu(\pi) = \max_{\pi \in \Pi} \mathbb{E}_{s_1 \sim \mu}[V_\mathcal{M}^\pi(s_1)] \tag{3}[2]$$
$$= \max_{\pi \in \Pi} \mathbb{E}_{\tau \sim T_\pi(\tau)}[R(\tau)] \tag{4}$$
where $T_\pi(\tau) = \mu(s_1) \prod_{t=1}^{\infty} \pi(a_t|s_t) P(s_{t+1}|s_t, a_t)$

With the basics of RL terminology and notation, we will review several leading RL algorithms in this section, with the focus on the mathematical formulation and foundational design. We will start with the model-free RL including value-based methods, policy gradient and actor-critic, and then briefly introduce model-based RL. A non-exhaustive taxonomy of RL algorithms can be found in Figure 1.

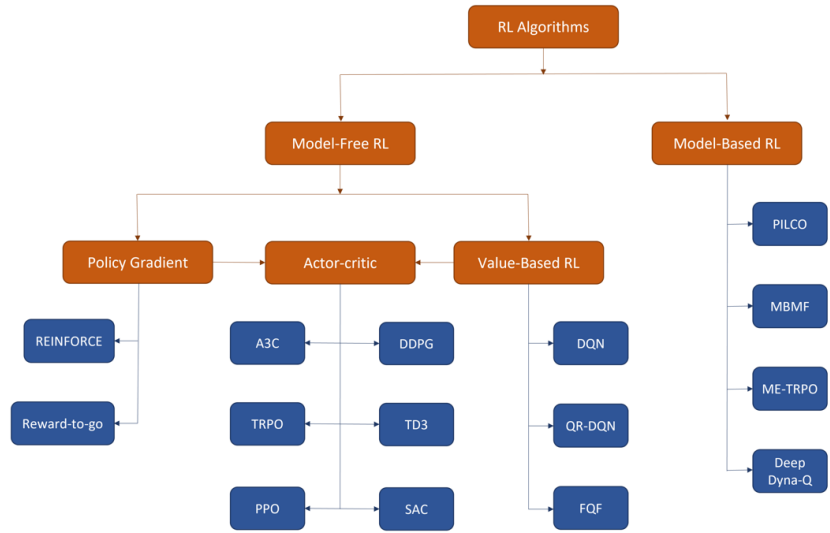

**Figure 1. A taxonomy of RL algorithms**

---

[2] Please note in the following subsections, we will drop the subscripts of $\mathcal{M}$ and $\mu$ from value functions and the objective function $J_\mu(\pi)$ respectively given working under the same MDP $\mathcal{M}$.

## Value-based Methods

A common way to optimize the RL objective $J(\pi)$ is by the observation that if the optimal value function or Q-function could be accurately estimated, we can easily recover an optimal policy. For instance, given the optimal Q-function $Q^*(s,a)$, an optimal policy $\pi^*(s)$ could be obtained by

$$\pi^*(s) = \text{argmax}_{a \in \mathcal{A}} Q^*(s,a) \tag{5}$$

Dynamic Programming (DP) is a classic approach to approximate these desired value functions assuming a perfect model of the environment is given, and the number of states and actions is small so that value functions can be represented in some lookup tables [18]. The foundation of DP is centered on bellman optimality equation and bellman expectation equation, and they can be defined as follows with respect to Q-function:

$$Q^*(s,a) = r(s,a) + \gamma \mathbb{E}_{s' \sim P}[\max_{a'} Q^*(s',a')] \tag{6}$$
$$Q^\pi(s,a) = r(s,a) + \gamma \mathbb{E}_{a' \sim \pi, s' \sim P}[Q^\pi(s',a')] \tag{7}$$

Where $s'$ is the successor state and $a'$ is the action performed at $s'$.

According to these equations, two helpful mathematical operations, bellman optimality operator $\mathcal{T}: \mathbb{R}^{|\mathcal{S}||\mathcal{A}|} \to \mathbb{R}^{|\mathcal{S}||\mathcal{A}|}$ and bellman evaluation operator $\mathcal{T}^\pi: \mathbb{R}^{|\mathcal{S}||\mathcal{A}|} \to \mathbb{R}^{|\mathcal{S}||\mathcal{A}|}$ can be constructed as follows:

$$(\mathcal{T}Q)(s,a) := r(s,a) + \gamma \mathbb{E}_{s' \sim P}[\max_{a'} Q(s',a')] \tag{8}$$
$$(\mathcal{T}^\pi Q)(s,a) := r(s,a) + \gamma \mathbb{E}_{a' \sim \pi, s' \sim P}[Q(s',a')] \tag{9}$$

These operators map any Q-function from $\mathbb{R}^{|\mathcal{S}||\mathcal{A}|}$ to another Q-function in the same space. It's helpful to consider the Q-function as a vector of length $|\mathcal{S}||\mathcal{A}|$ and the operators are just some transformations that take the vector and output another vector with the same dimensions. More specifically, considering $\mathcal{T}$ and a state-action value $Q(s,a)$, $\mathcal{T}$ can be viewed as an assignment statement that updates the original value of $Q(s,a)$ with the one derived from the RHS of (8). There are two pleasant properties of these operators, related to contraction mapping, which help the design of the classic DP algorithms. First, given any arbitrary Q-function $Q$, repeatedly applying $\mathcal{T}^\pi$ or $\mathcal{T}$ on $Q$ yields $Q^\pi$ or $Q^*$ respectively. This property can directly turn Bellman operators into update rules, providing an iterative algorithm for approximating a desired Q-function. Moreover, $\mathcal{T}^\pi$ and $\mathcal{T}$ have unique fixed points $Q^\pi$ and $Q^*$ such that $\mathcal{T}^\pi Q^\pi = Q^\pi$ and $\mathcal{T}Q^* = Q^*$. This second property serves as the termination condition of many classic DP algorithms and the building blocks of some advanced RL algorithms, such as actor-critic.

Since, in practice, the complete knowledge of the environment is usually unknown, and the number of states and actions can be arbitrarily large, dynamic programming is thus limited to some restricted problems by its assumptions and function representation. The value-based RL, sometimes known as approximate dynamic programming, provides a class of algorithms that overcome these problems, extending the framework of iterative dynamic programming with modified bellman operators and function approximation. The essence of this approach is to modify the update rule by approximating the bellman operators with some empirical estimators, i.e., estimating the RHS of equations (8) and (9) with sampling [19]. In the simplest case of using a single sample $<s, a, r, s'>$ where $r = r(s, a)$, the empirical bellman operators $\hat{\mathcal{T}}Q$ and $\widehat{\mathcal{T}^\pi}Q$ can be constructed as [20–22]:

$$(\hat{\mathcal{T}}Q)(s, a) := r(s, a) + \gamma \max_{a'} Q(s', a') \tag{10}^3$$
$$(\widehat{\mathcal{T}^\pi}Q)(s, a) := r(s, a) + \gamma Q(s', \pi(s')) \tag{11}$$

A classic algorithm under this category is the Fitted Q-Iteration (FQI) [23]. In FQI, the algorithm first gathers a dataset $D = \{<s_i, a_i, r_i, s_{i+1}>\}_i^{|D|}$ where state $s_i$ and action $a_i$ are drawn from a pre-defined distribution $\nu$, and the next state $s_{i+1}$ and the reward $r_i$, are obtained from an unknown state transition probability function $P(\cdot | s_i, a_i)$ and reward function $r(s_i, a_i)$ of the environment. With the initialization of a parameterized Q-function $Q_\theta$, the Q-values $Q_\theta(s_i, a_i)$ are updated towards their target $\hat{\mathcal{T}}Q_\theta(s_i, a_i)$ and the update rule can be formulated as

$$\theta \leftarrow \text{minimize}_\theta \sum_{i=1}^{|D|} \left( \hat{\mathcal{T}} Q_\theta(s_i, a_i) - Q_\theta(s_i, a_i) \right)^2 \tag{12}$$
Where $\hat{\mathcal{T}} Q_\theta(s_i, a_i) = r_i + \gamma \max_{a_{i+1}} Q_\theta(s_{i+1}, a_{i+1})$

Equation (12) can be interpreted as applying Monte-Carlo approximation of Bellman optimality $\hat{\mathcal{T}}$ to Q-function $Q_\theta$ through minimizing the square loss over $\theta$. And since the Q-values in FQI are estimated by a parameterized function rather than a lookup table (as in DP), they are closely correlated to each other, implying that a small update of $\theta$ may benefit some Q-values but push others away from their targets. Therefore, unlike supervised learning where the ground truth labels remain unchanged, the targets $\hat{\mathcal{T}}Q$ defined in FQI may vary every time when the parameter changes, which introduces instability in training. Consequently, depending on the generalization and extrapolation ability of the function approximation, the contraction mapping property of DP cannot guarantee convergence under parametrized Q-functions, especially for those using neural networks [24]. However, there exist other variants of FQI using non-parametric approximation architecture, such as k-nearest-neighbor and totally randomized trees, showing strong convergence property [25–28].

Regarding the value-based methods using neural networks, Deep Q-Network algorithm, showing strong performance in a variety of ATARI games, can be viewed as an instantiation of FQI in online settings [29]. The Q-function approximator of DQN can be any typical neural network, referred to as Q-network. The framework of DQN uses two heuristics to limit the instability inherited from FQI with neural networks. First, a separated network called target network is introduced solely for computing the targets (interpreted as the ground truth) due to their inconsistency. Compared to Q-network, the target network is updated less frequently to keep the target fixed for an extra amount of time. With this strategy, DQN prevents the instability from propagating too quickly and thus reduces the risk of divergence.

Moreover, Deep Q-Network is an online learning algorithm which follows its current policy for exploring and data sampling. Without proper care, this may deliver a policy that has inferior performance since previously visited state-action pairs with high rewards may not guarantee to be revisited due to the stochasticity of the system. Experience Replay, a replay buffer, is introduced to solve this issue by keeping all samples $<s, a, r, s'>$ from the last $N$ steps in the memory where $N$ is usually very large. Besides, ER selects samples from the memory following the uniform distribution for every mini-batch update. This sampling method, though simple, effectively breaks the temporal correlation between samples in the same trajectory, which helps reduce bias. With large memory and uniform sampling, mini-batch samples constructed under ER are more representative than those from alternative methods such as Q-learning [30] which uses only one sample for an update. Additionally, a large buffer provides good coverage of the state-action space, which makes the policy training more exploratory and consistent, thus increasing the chance of achieving a more desirable policy.

---
[3] It is worth noting that the empirical bellman operator constructed in (10) is an analogue to Q-learning [30] with the learning rate $\alpha = 1$. Theoretically, if we want to turn (10) directly into an update rule, a large number of successor states $\{s_i'\}_i^n$ are required for convergence, i.e., $(\hat{\mathcal{T}}Q)(s, a) := r(s, a) + \gamma \sum_i^n \max_{a'} Q(s_i', a')$ [21]. Otherwise, a new algorithm with a customized update rule needs to be proposed [22].

In addition to target network and experience replay, there are other advanced approaches that could help improve the stability and efficiency of policy learning including Double DQN [31] for overestimation reduction, Multi-step learning for accurate target estimation [24], Dueling Networks [32] for enhanced Q-function representation and Prioritized Experience Replay (PER) for efficient TD-error minimization [33], along with other extensions such as Distributional DQN [34], Quantile Regression DQN (QR-DQN) [35], and Fully parameterized Quantile Function (FQF) [36].

Policy gradient

Unlike the value-based approach that learns a desired policy by estimating the optimal value functions, the Policy Gradient (PG) methods work directly on the objective function defined in (4) or some equivalent ones. In the setting of this review, we limit our focus on (4). Specifically, with a policy $\pi$ parameterized by $\theta \in \Theta \subset \mathbb{R}^d$, the goal of PG is to find a $\theta$ that maximizes $J(\pi_\theta)$:

$$\max_{\theta \in \Theta} J(\pi_\theta) \tag{13}$$

Since the search space has now shifted from the policy space $\Pi$ in (4) to $\Theta$ which is continuous and has fixed dimensions, various numerical optimization methods could be utilized to solve (13). Gradient ascent is one direct approach to this, which iteratively moves in the direction of steepest ascent as defined by the gradient for maximizing $J(\pi_\theta)$. Following is the expression of the gradient for $J(\pi_\theta)$:

$$\nabla_\theta J(\pi_\theta) = \nabla_\theta \mathbb{E}_{\tau \sim T_{\pi_\theta}(\tau)}[R(\tau)]$$

$$= \mathbb{E}_{\tau \sim T_{\pi_\theta}(\tau)} \left[ \sum_{t=1}^{\infty} \nabla_\theta \log \pi_\theta(a_t|s_t) R(\tau) \right] \tag{14}$$

Where $R(\tau) = \sum_{t=1}^{\infty} \gamma^{t-1} r(s_t, a_t)$.

At each update iteration $k$, gradient ascent, with a fixed step size $\alpha$, follows the update rule:

$$\theta_{k+1} = \theta_k + \alpha \nabla_{\theta_k} J(\pi_{\theta_k}) \tag{15}$$

Since the policy gradient $\nabla_\theta J(\pi_\theta)$ is an expectation over all possible trajectories, it can be directly estimated by a set of samples where each is a trajectory collected by the agent interacting with the environment while following policy $\pi_\theta$. With the set of samples denoted as $D$ and the length of trajectories assumed to be $H$, we can estimate $\nabla_\theta J(\theta)$ as:

$$\nabla_\theta J(\theta) \approx \frac{1}{|D|} \sum_{\tau \in D} \nabla_\theta \log T_{\pi\theta}(\tau)[R(\tau)] \tag{16}$$

$$\approx \frac{1}{|D|} \sum_{i=1}^{|D|} \sum_{t=1}^{H} \nabla_\theta \log \pi_\theta(a_{i,t}|s_{i,t}) R(\tau) \tag{17}$$

Where $R(\tau) = \sum_{t=1}^{H} \gamma^{t-1} r(s_t, a_t)$.

With equation (17), we have derived our first policy gradient method, commonly known as REINFORCE [37] which serves as the foundation of PG-based methods.

Because $\nabla_\theta J(\pi_\theta)$ is an expectation, a long-standing issue centered on PG-based methods is to derive an unbiased estimator of $\nabla_\theta J(\pi_\theta)$ with possibly low variance. For instance, $R(\tau)$ in (17) can be replaced by a smaller term $\sum_{t'=t}^{H} \gamma^{t'-t} r(s_{t'}, a_{t'})$ thanks to causality that the policy at time $t'$ cannot affect the reward at time $t$ when $t < t'$. This gives us a new unbiased estimator of $\nabla_\theta J(\pi_\theta)$ with reduced variance called Reward-to-go [38]:

$$\nabla_\theta J(\theta) \approx \frac{1}{|D|} \sum_{i=1}^{|D|} \sum_{t=1}^{H} \gamma^{t-1} \nabla_\theta \log \pi_\theta(a_{i,t}|s_{i,t}) \left( \sum_{t'=t}^{H} \gamma^{t'-t} r(s_{i,t'}, a_{i,t'}) \right) \tag{18}$$

Regarding the policy gradient $\nabla_\theta J(\pi_\theta)$, it can be viewed as an analogue of the gradient in maximum likelihood where the term $\log \pi_\theta(a_{i,t}|s_{i,t})$ in equations (17) and (18) is the log likelihood of the policy $\pi_\theta$ given data $<s_{i,t}, a_{i,t}>$. Unlike maximum likelihood, the log likelihood in PG is now weighted by some discounted cumulative rewards, such as $R(\tau)$ or $\sum_{t'=t}^{H} \gamma^{t'-t} r(s_{i,t'}, a_{i,t'})$. Thus, updating the parameter $\theta$ with $\nabla_\theta J(\pi_\theta)$ would change the policy according to the sign of $R(\tau)$ or $\sum_{t'=t}^{H} \gamma^{t'-t} r(s_{i,t'}, a_{i,t'})$, e.g., the chance of selecting $a_{i,t}$ given $s_{i,t}$ increases if $R(\tau)$ is positive or decreases otherwise.

On the other hand, if the reward function only outputs positive values and some mediocre actions are randomly selected for parameter updates, the agent may tend to choose these inferior actions even their absolute values of $\log \pi_\theta(a_{i,t}|s_{i,t})$ are small. A popular strategy to address this unstable issue is to introduce an extra term, called baseline $b$, which can be shown to always keep the estimator of $\nabla_\theta J(\theta)$ unbiased while reducing its variance if selected properly [39]. For instance, by choosing an average reward over sampled trajectories as baseline, the term $\left( \sum_{t'=t}^{H} \gamma^{t'-t} r(s_{i,t'}, a_{i,t'}) \right) - b$ in (19) is centered around 0, which improves the sampling efficiency by distinguishing actions with their relative rewards.

$$\nabla_\theta J(\theta) \approx \frac{1}{|D|} \sum_{i=1}^{|D|} \sum_{t=1}^{H} \gamma^{t-1} \nabla_\theta \log \pi_\theta(a_{i,t}|s_{i,t}) \left[ \left( \sum_{t'=t}^{H} \gamma^{t'-t} r(s_{i,t'}, a_{i,t'}) \right) - b \right] \tag{19}$$

Where $b$ can be $\frac{1}{|D|} \sum_{i=1}^{|D|} R(\tau)$

Thus, with proper care of designing the baseline, the policy $\pi_\theta$ could eliminate the inferior actions and choose more desirable ones with higher probability. In the next subsection, a learnable baseline will be introduced, which is more stable and significantly reduces the variance of the classic policy gradient estimators.

Actor-Critic

Actor-critic algorithms can be viewed as an approach that combines policy gradients with value-based methods to optimize J(θ). Generally, it takes a policy $\pi_\theta$ as an actor to interact with the environment, while maintaining a learnable value function as the critic to evaluate the actor's actions [24, 40]. The simplest form of actor-critic called Q actor-critic [40] can be derived directly from the reward-to-go and we denote $\widehat{\nabla_\theta J(\theta)}$ below as the estimator of $\nabla_\theta J(\theta)$ from (18):

$$\mathbb{E}\big[\widehat{\nabla_\theta J(\theta)}\big] = \mathbb{E}\left[\frac{1}{|D|}\sum_{i=1}^{|D|}\sum_{t=1}^{H}\gamma^{t-1}\widehat{Q^{\pi_\theta}}(s_t, a_t)\nabla_\theta \log \pi_\theta(a_{i,t}|s_{i,t})\right]$$

$$= \mathbb{E}\left[\frac{1}{|D|}\sum_{i=1}^{|D|}\sum_{t=1}^{H}\gamma^{t-1}\mathbb{E}\big[\widehat{Q^{\pi_\theta}}(s_t, a_t)\big|s_t, a_t\big]\nabla_\theta \log \pi_\theta(a_{i,t}|s_{i,t})\right]$$

$$= \mathbb{E}\left[\frac{1}{|D|}\sum_{i=1}^{|D|}\sum_{t=1}^{H}\gamma^{t-1}Q^{\pi_\theta}(s_t, a_t)\nabla_\theta \log \pi_\theta(a_{i,t}|s_{i,t})\right] \tag{20}$$

Where $\widehat{Q^{\pi_\theta}}(s_t, a_t) = \sum_{t'=t}^{H}\gamma^{t'-t}r(s_{i,t'}, a_{i,t'})$

A direct advantage of this new estimator, the one inside the expectation of (20), is that it has less variance than the reward-to-go. This can be shown following the tower property of conditional probability and the definition of variance. The intuition is that $\widehat{Q^{\pi_\theta}}(s_t, a_t)$ is indeed the cumulative reward of a single sample trajectory that estimates $Q^{\pi_\theta}(s_t, a_t)$, which inevitably has larger variance and so does the reward-to-go. In practice, we usually estimate $Q^{\pi_\theta}$ by a parameterized Q-function $Q^{\pi_\theta}_\omega$ and update $\omega$ following a similar procedure as Fitted Q-Iteration described in the previous subsection of value-based methods.

Similar to (19), a baseline can be incorporated into Q actor-critic to further reduce its variance. A favorable choice is using a value function $V^{\pi_\theta}_\varphi$ parameterized by $\varphi$ where $\nabla_\theta J(\pi_\theta)$ can be further reduced to a form of advantage function:

$$\nabla_\theta J(\theta) \approx \frac{1}{|D|}\sum_{i=1}^{|D|}\sum_{t=1}^{H}\gamma^{t-1}\nabla_\theta \log \pi_\theta(a_{i,t}|s_{i,t}) A^{\pi_\theta}_{\omega,\varphi}(s_{i,t}, a_{i,t}) \tag{21}$$

Where $A^{\pi_\theta}_{\omega,\varphi}(s_{i,t}, a_{i,t}) = Q^{\pi_\theta}_\omega(s_{i,t}, a_{i,t}) - V^{\pi_\theta}_\varphi(s_{i,t})$

The derived PG estimator in (21) is the naïve advantage actor-critic [24, 41, 42]. As the advantage function measures the performance difference between a specific action and the average action in a given state, it's usually considered as a favorable critic to evaluate the actor and determine which action should be chosen more often.

Moreover, $\nabla_\theta J(\pi_\theta)$ can be approximated in another way as specified in (22) by the observation that given a sample $<s, a, r, s'>$, the sum of $r$ and $\gamma V^\pi(s')$ is a sample estimate of $Q^\pi(s,a)$:

$$\nabla_\theta J(\theta) \approx \frac{1}{|D|}\sum_{i=1}^{|D|}\sum_{t=1}^{H}\gamma^{t-1}\nabla_\theta \log \pi_\theta(a_{i,t}|s_{i,t})A_\varphi^{\pi_\theta}(s_{i,t}, a_{i,t}) \quad (22)$$
Where $A_\varphi^{\pi_\theta}(s_{i,t}, a_{i,t}) = r(s_{i,t}, a_{i,t}) + \gamma V_\varphi^{\pi_\theta}(s_{i,t+1}) - V_\varphi^{\pi_\theta}(s_{i,t})$

The unbiased estimator of $\nabla_\theta J(\pi_\theta)$ derived in (22) is called TD-error actor-critic [41, 43, 44]. Theoretically, it has a higher variance than the naïve advantage actor-critic in (21) as we replace $Q_\omega^\pi(s_{i,t}, a_{i,t})$ with a single sample estimate $r(s_{i,t}, a_{i,t}) + \gamma V_\varphi^{\pi_\theta}(s_{i,t+1})$. But in practice the estimator of $\nabla_\theta J(\pi_\theta)$ in (22) is more stable for policy training and has less bias as it only requires a function estimator for the value function $V_\varphi^{\pi_\theta}$ while the naïve advantage actor-critic requires an extra function estimator $Q_\omega^{\pi_\theta}$ for Q-function other than $V_\varphi^{\pi_\theta}$, which thus needs more samples to achieve a comparable performance. Owing to the favorable property of TD-error actor-critic, precisely the term of $A_\varphi^{\pi_\theta}(s_{i,t}, a_{i,t})$, it has been widely adapted to many advanced actor-critic algorithms such as Asynchronous Advantage Actor Critic (A3C) [42], Trust Region Policy Optimization (TRPO) [45] and Proximal Policy Optimization (PPO) [46].

The methods discussed so far in this subsection, as well as the ones mentioned in PG, are generally on-policy methods where the target policy being learned (i.e., evaluated and improved) is used as the one to interact with the environment and generate samples, which is commonly referred to as behavior policy [24]. On the other hand, in off-policy setting, agent usually learns a policy that is different from the one being executing. Some typical examples are the methods introduced in the value-based RL, such as FQI, Q-learning and DQN, where the samples used for training are not exactly generated by the current learning policy, the target policy. There is a class of off-policy actor-critic methods, such as Deep Deterministic Policy Gradient (DDPG) [47], Twin Delayed DDPG (TD3) [48] and Soft Actor-Critic (SAC) [49], that adopt the replay buffer and heavily depend on Q-function estimator for approximating desired policy. These methods, in practice, are usually more sample efficient yet less robust to hyperparameter settings than their on-policy counterparts mainly due to the incorporation of off-policy samples and sometimes convoluted tricks for stabilizing training.

Model-based RL

Model-based RL is a general term that refers to a broad class of algorithms, which is widely seen as a potential approach to improve the sample efficiency of model-free RL [50–52]. Unlike model-free RL which learns a policy $\pi_\theta$ and/or a value function, without modeling the environment, model-based RL explicitly estimates the transition dynamics $P(s_{t+1}|s_t, a_t)$, which we denote $P_\psi(s_{t+1}|s_t, a_t,)$ parameterized by $\psi$, and uses it for planning [53] or policy learning. Since model-based RL has not been well standardized and there exist many variants, we will briefly summarize three important types of algorithms. A common class of model-based RL algorithms, related to shooting methods [54], learns the dynamics model $P_\psi$, and uses it to derive a local planning strategy for action selection. This is typically achieved by formulating and solving receding horizon problems posed in model-predictive control [55] using various trajectory optimization methods [56, 57]. Other model-based RL methods, such as PILCO, learn a policy $\pi_\theta$ in addition to the dynamics model $P_\psi$, and employ backpropagation through time with respect to the expected future reward for policy search and improvement [58–60]. Dyna-style algorithms, another set of model-based methods, incorporate the dynamic model $P_\psi$ to the general model-free framework for augmenting samples and accelerating policy learning [61–64]. Specifically, training under this category usually iterates between two steps: first, the agent learns $P_\psi$ and $\pi_\theta$ with a set of real experiences $\{<s_i, a_i, r_i, s_{i+1}>\}_i^n$ gathered from interacting with the environment. Second, a number of 'synthetic' samples $\{<\hat{s}_i, \hat{a}_i, \hat{r}_i, \hat{s}_{i+1}>\}_i^m$ are generated using the current policy $\pi_\theta$ under the learned dynamic model $P_\psi$ and another round of policy update is performed with these 'synthetic' samples. Such an iterative process can significantly boost the sample efficiency of pure model-free learning once $P_\psi$ precisely estimates the environmental dynamics.

## 3. State-of-the-arts in applying reinforcement learning to drug discovery

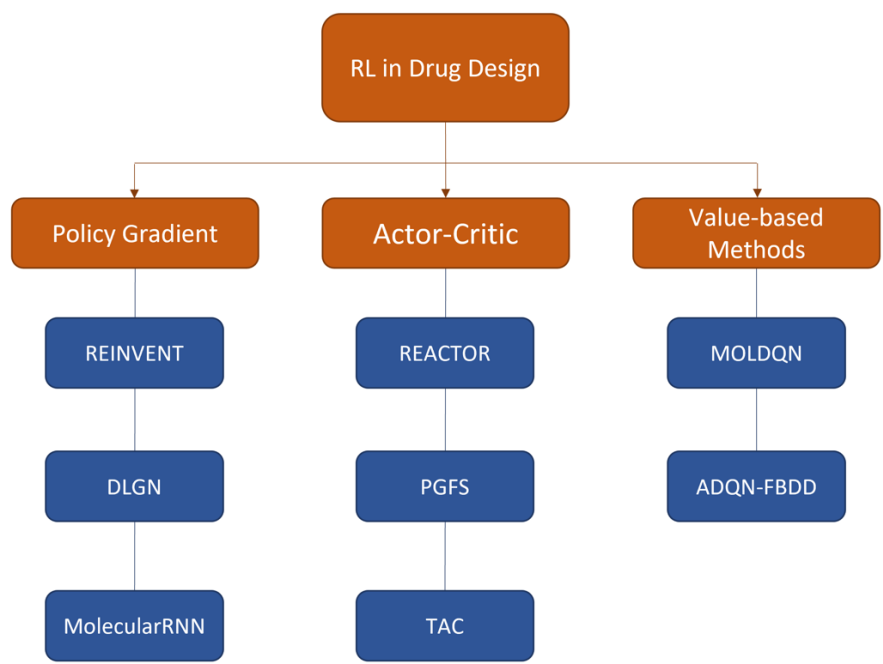

**Figure 2**. RL-based methods for drug design

There are different approaches in applying reinforcement learning to drug discovery depending on the objectives [65–67]. Distributional learning, a common task of computational drug design, is to generate a set of molecules distributed differently from an existing dataset, which satisfy one or more predefined requirements. One typical approach is to pretrain a generative model with some well-established dataset and incorporate it into a RL framework for optimization. In this case, the generative model is trained as learning a policy that maximizes a RL loss function as defined in (3), where the reward is customized for the objective(s). Another set of problems, usually referred to as goal-directed learning, requires searching for the exact molecule with specific properties.

These tasks are usually formulated as solving a combinatorial optimization problem [68, 69], which falls exactly into the region that could be potentially solved by RL [70, 71]. In this section, we will briefly review different adaptations of RL algorithms, as defined in section 2, to drug discovery, with a focus on the practical strategy and design of RL-based generative models. A non-exhaustive taxonomy of RL-based methods for drug design can be found in Figure 2.

### Models with policy gradient

Policy gradient method has been adapted to a variety of RL-based generative models for distributional learning owing to its policy stochasticity and learning capability for high-dimensional state and/or action spaces. Marcus et al. developed an approach REINVENT for distributional learning based on REINFORCE algorithm, in which they first pre-trained a recurrent neural network (RNN) known as prior policy that could generate a set of samples with a similar structural distribution as ChEMBL [72]. Then the agent RNN for learning target policy was initialized with the same architecture and parameters as the pre-trained RNN and later tuned using RL to achieve higher expected score while keeping the new policy close to the prior policy. The generated molecules thus have a similar structural distribution as the ChEMBL dataset while being optimized towards the target properties. Since the agent RNN is pre-trained, the RL searching efficiency has been increased whereas the exploration capability of this model is somewhat limited.

Recently, the attention to polypharmacology is constantly increasing owing to its therapeutic potential in some complex pathologies. Dual-target ligand generative network (DLGN), leveraging RL and adversarial training, was developed to generate molecules that have bioactivities toward two targets [73]. With SMILES string as input, DLGN uses an RNN-based generator to produce novel molecules that satisfy the predefined constraints. To make generated molecules dual-targeted, DLGN utilizes two discriminators to monitor the generative process and encourages the generated molecules lying in the intersection of the two bioactive-compound distributions. The drawback of this model is that, although the generator does not need to be trained with labeled data, the discriminators require reliable labeled data to control the qualities of generated molecules.

As graph is a more natural way to represent a molecule, graph neural networks (GNNs) are widely used in computational drug discovery. You et al. developed a graph convolutional policy network (GCPN) that generates molecules under some guidance towards desired objectives while restricting the output molecules to some specific chemical rules [8]. To achieve this goal, they leveraged RL for molecule generation and optimization under a simulated environment with a designed reward function, and used expert pre-training and adversarial training to incorporate prior knowledge for guidance. In another work, Sara et al. extended REINVENT with gated graph neural networks to generate molecules with desired properties [74]. To overcome the difficulty of RL training, they introduced a best agent reminder (BAR) loss, which is calculated based on the actions given by the current agent and the best agent, and shown to significantly improve the speed of convergence and the final performance. MolecularRNN is another interesting work based on the REINFORCE algorithm that utilizes graph data structure to represent molecules while exploiting RNN as the generator. It employs a dual network model, consisting of both node-level and edge-level RNNs, to predict the next atom and bond given an intermediate generated molecular graph [75]. With valency-based rejection sampling constraints, MolecularRNN achieves 100% validity of the generated molecules in the experiments.

To balance the exploration and exploitation, Xuhan et al. employed two pretrained RNNs in DrugEx to generate novel molecules [76]. The two RNNs share the same architecture while having different internal parameters. One of the RNN, which serves as the exploration network, is pretrained on the ZINC database and then fine-tuned with desired molecules to memorize the distribution of the potential drug space. This exploration network will not be updated during training. The other RNN, performing as exploitation network, is only pretrained on ZINC database, followed by policy gradient update to generate molecules with high pIC50 values for properties of interests. During the training phase, the exploration and the exploitation networks are randomly assigned to generate the next token based on a predefined 'exploring rate'. This helps the generated molecules to be well diverse while maintaining high pIC50 value for target proteins. To improve the balance of exploration and exploitation, Tiago et al. extended DrugEx by introducing a dynamically adaptive 'exploring rate' which is determined by the property of two latest batches of generated molecules [77]. Additionally, they added a penalty in reward for improving novelty when the diversity of the latest generated molecules decreases.

### Models with Deep Q-networks (DQNs)

Value-based RL methods are usually more stable and sample efficient than those using policy gradient [24]. In addition, as the policy learned by value-based approach is deterministic, it's more natural to choose these methods for solving the goal-directed problems. Molecule Deep Q-Network (MolDQN) based on bootstrapped-DQN is a leading method for goal-directed learning [78]. By allowing only valid actions, MolDQN guarantees that the generated molecules are 100 % valid. The molecule generation or optimization starts from an empty or a seed molecule in the form of Morgan fingerprint. In the constrained optimization task, MolDQN optimizes the seed molecule toward the target property while maintaining its similarity to the original molecule above a designed threshold. The advantage of MolDQN is that it does not depend on a pre-trained model for molecule generation, and thus it is not biased to any observed chemical space. Therefore, MolDNQ can in principle generate novel chemical structures or molecules with desired properties but may require considerable time for exploring to achieve favorable performance.

Tang et al. developed an advanced deep Q-learning network with the fragment-based drug design (ADQN-FBDD) to generate molecules specifically targeting a protein with known 3D structure [79]. They designed a practical reward by considering drug-likeness, as well as specific fragments and pharmacophores which are protein-structure dependent. Although this approach has a obvious limitation, which is the high dependency of 3D structures of target proteins, a recent breakthrough in protein structure prediction [15] may provide it with a wide range of application scenarios.

### Models with actor-critic

As mentioned in section 2, the actor-critic method improves the sampling efficiency of policy gradient by learning a parameterized value function (critic). It usually involves two networks, a policy network and a variance-reduced value network. Intuitively, the actor, the policy network, decides what action to take based on the current policy and state whereas the critic, value network, evaluates the action and informs the actor how the policy should be adjusted. With the guidance from the critic, the policy training process usually becomes more stable and efficient [80].

Gregor et al. developed a novel actor-critic architecture for 3D molecule design that exploits the symmetries of the design process [81]. Specifically, it employs a state embedding network to obtain rotation-covariant and -invariant state representations of 3D molecular graphs, which improves the generalization of the policy network and enables it to generate more complex 3D structure than previous approaches. Niclas et al. introduced a fragment-based RL framework based on actor-critic where both actor and critic are modeled with bidirectional long short-term memory (LSTM) networks [82]. As the research is focused more on the constrained optimization task and the molecular fragments are used as atomic actions, a novel encoding approach using a balanced binary tree was developed to represent the fragments. With the help of this design, the RL agent is more capable of distinguishing promising steps from those mediocre ones.

Generating molecules in silicon is far from the final goal of developing drugs. The first step after molecular design is to synthesize designed compounds and test their effects with wet experiments. However, not all computational designed molecules are synthesizable. This seriously affects the practical value of computational drug design. Some studies try to resolve this by taking synthesizability into consideration while generating molecules. Reaction-driven objective reinforcement (REACTOR) empowered by actor-critic method, for example, defines the state-action trajectory in RL as a sequence of chemical reactions, and thus not only improves the synthesizability of generated molecules, but also speeds up the exploration rate of the model in the chemical space [83]. Additionally, REACTOR employs a synchronous version of A3C which can perform parallelized policy search and thus tremendously improves the efficiency of the policy training. In addition to REACTOR, there are other works that leverage actor-critic methods for synthesis-oriented molecule generation, such as Towered Actor-Critic (TAC) [84] and Policy Gradient for Forward Synthesis (PGFS) [85].

Remarks on RL algorithms

With the jumbo chemical searching space and limited pretraining samples, machine learning algorithms like GAN and VAE have difficulties to generate a set of molecules distributing differently from the training dataset. Moreover, such methods are inherently hard to be adapted for goal-directed tasks, especially for those related to exhaustive search. Under such circumstances, RL, which does not rely on training data and can be adopted to skew the underlying distribution of the (pretrained) generative model or directly learn a policy to optimize a seed molecule, is a promising approach to tackle various drug discovery problems. However, the current RL methods applied for molecule generation are generally sample inefficient especially in high-dimensional search space under multiple constraints, which is usually the case for drug-like molecule space. Recent studies tend to incorporate pretraining and adversarial training into conventional RL framework to overcome this difficulty. Moreover, the quality of the molecules generated by RL-based generative model highly relies on the reward function. Thus, in future studies, improving the sample efficiency of RL and the accuracy of the property predicting model is the key to bring computational drug design to practical production.

# 4. Challenges and opportunities of reinforcement learning in systems pharmacology and personalized medicine

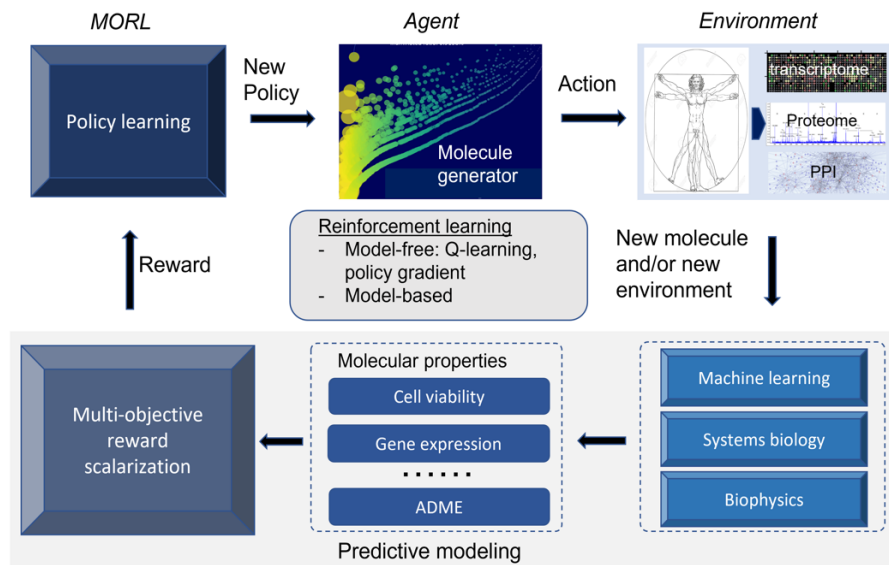

**Figure 3**. Illustration of a RL framework for systems pharmacology-oriented lead optimization.

Figure 3 illustrates a RL framework for systems pharmacology-oriented personalized lead optimization and drug design. The molecule and the environment together constitute a state. A molecule generator (agent) will first take an action based on the current state and policy to generate a new molecule or modify a seed molecule (e.g., replacing a hydrogen atom with a methyl group). Then, multiplex phenotypic responses (cell viability, drug-target profile, chemical-induced gene expression, pharmacokinetics etc.) in an individual patient (environment) will be predicted for the newly generated molecule by machine learning, biophysics, systems biology, or other methods, and these responses are used as the reward for policy training. A new policy will be learned based on observed actions, states, and rewards by performing an optimization with a multi-objective RL (MORL) algorithm, and a new molecule will be generated again from the updated policy. Unlike target-based compound screening where only chemical space is needed to be explored, systems pharmacology-oriented personalized drug discovery needs to optimize the interplay of chemicals, the druggable genome, and high-dimensional omics characterizations of disease models or patients (environment). Several barriers need to be overcome when applying RL to systems pharmacology and precision medicine. These include the exploration of out-of-distribution samples, generalization power of RL, adaptive multi-objective optimization [86–88], and activity cliffs of quantitative structure-activity relationship (QSAR) space [89].

*Out-of-distribution reward function.* Although rewards for several molecular properties (e.g., logP and druglikeness) can be directly calculated from a given molecular structure, reward functions that are the most important for drug actions such as binding affinity and chemical-induced gene expression need to be obtained from a predictive model that is dependent on machine learning, mathematical or physics-based modeling. In spite of tremendous advances in deep learning and availability of diverse omics data sets, robust and accurate predictions of genome-wide drug-target interactions and molecular phenotypic read-outs in a physiological relevant condition remain as an unsolved challenging problem. The problem is rooted in biased, noisy, and incomplete omics data and inherited from the limitation of machine learning. In the case of genome-wide drug-target prediction, only less than 10% of gene families have known small molecule ligands. The remaining 90% gene families are dark matters in the chemical genomics space [17]. Even for mostly studied G-protein coupled receptors (GPCRs), more than 99% receptors are orphans, i.e., their endogenous or exogenous ligands are unknown. Similarly, only a small number of cell lines have annotated drug response data. It is a fundamental challenge to generalize a "well-trained" machine learning model to unseen data (e.g., patients), which lie out-of-the-distribution (OOD) of the training data (e.g., cell lines), so as to successfully predict outcomes from conditions that the model has never before encountered. While deep learning is capable, in theory, of simulating any functional mapping, its generalization power is notoriously limited in the case of distribution shifts [90].

*Generalization power of RL.* The conventional MDP defined in section 2 is assumed that the problem setting is stationary, i.e., the transition dynamics and the reward function do not change over time, and the agent fully observes the underlying state, i.e., the observation received by the agent includes perfect information of the current state. Thus, the conventional RL methods, especially those falling into the model-free RL, may perform poorly in an environment that is non-stationary (e.g., from one patient to another) with partially observed state though they perform well in the conventional setting [91, 92]. Additionally, directly employing conventional RL methods to solve an MDP with corrupted[3] or sparse reward[4] could fail catastrophically [93, 94]. The partially observed, non-stationary or corrupted sparse-reward environment is the exact situation in systems pharmacology. Due to observed chemical activity and omics data highly biased and incomplete, it is likely that a novel molecule is an OOD sample. Thus, no reliable reward can be assigned to this molecule as mentioned above. Additionally, drug response data mainly comes from cell lines or disease models. The environment in a cell line or animal model could be dramatically different from human bodies. A naïve adaptation of standard RL systems is not sufficient for systems pharmacology. New methods are needed to improve the robustness, generalizability and transferability of RL.

*Adaptive multi-objective optimization*. To design drugs that optimize the system-level responses to maximize therapeutic effects and minimize side effects, it is needed for a RL algorithm to optimize multiple (sometimes conflict) objectives such as pharmacokinetics, blood-brain barrier permeation, drug binding affinities to multiple targets, or chemical-induced gene expression profiles. Although RL methods have been developed for multiple objectives [78], the final reward function in these methods is an linear combination of the reward functions of individual objectives with *a priori* defined weights. It is often difficult to define such weights. Moreover, the weight may be altered when an environment changes. For example, gene expression profiles can be dramatically different for different patients and in different disease states. Thus, for conventional RL, typically with fixed weight, a generative model (the policy) needs to be trained for different conditions. In addition, it is more computationally challenging to find optimal solutions in the framework of Pareto optimization for the multi-objective drug design due to the high dimensionality and uncertainty of omics and bioassay data. It has been suggested that Pareto optimization should be integrated with evolutionary algorithms or other complexity reduction methods [86, 95].

---

[3] In the corrupted reward problem, observed reward may not be an unbiased estimate of the true reward (e.g., the response data from cell lines is different than that of human bodies)

[4] In a dense-reward setting, almost every state-action pair in a trajectory is assigned with a reward. On the contrary, if rewards are not available for most of the state-action pairs, the setting is considered as sparse-reward.

*Activity cliff of QSAR.* Reward drop is a phenomenon that the reward $r_t$ received from a trajectory $\{s_1, a_1, s_2, a_2, ..., s_t, a_t\}$ suddenly drops or oscillates dramatically within certain range due to the nonsmoothness of the reward function, and many advanced RL algorithms have suffered from this problem [96]. Unfortunately, in computational molecule design, a well-known phenomenon in QSAR is activity cliff, in which a slight modification of chemical structure may lead to a dramatic activity change. The activity cliff is more complicated in systems pharmacology than single-targeted drug discovery. For example, the replace of a methyl group with an ethyl group for a chemical compound that has moderate binding affinities to two targets may increase the binding affinity to one of targets, but completely destroy the binding to another target due to steric clash. As a result, the phenotype response modulated by this compound could be changed significantly. Such unpleasant property (of QSAR landscape) thus increases the difficulty of policy learning for RL-based drug design.

To address aforementioned challenges in adapting RL to systems pharmacology and precision medicine, a synergistic integration of latest advances in RL with new development in machine learning and other related fields is needed.

*Improving robustness, generalizability, and transferability of RL algorithms.* Recently, Ghosh et al, showed that optimal generalization of RL at test-time corresponds to solving a partially-observed Markov decision process (POMDP) that is induced by the agent's epistemic uncertainty about the test environment [92]. They proposed an algorithm, LEEP, which uses an ensemble of policies to approximately learn the Bayes-optimal policy for maximizing test-time performance. In another study, Agarwal et al. proposed meta-reward learning to achieve the generalizability of RL in a sparse-reward environment [97]. In principle, these techniques can be adapted for systems pharmacology.

The majority of existing work in RL for drug design is based on the model-free approach. Model-based RL [6] may provide new opportunities for addressing the challenges in systems pharmacology-oriented drug discovery for precision medicine. Different from model-free approach, model-based method typically employs a learned dynamics model to facilitate policy training. It is more powerful in predicting future reward, has higher sample efficiency, and bears stronger transferability and generality than the model-free approach. The ability of predicting future reward may help to avoid activity cliffs. Sample efficiency will be helpful to alleviate issues in data sparsity, biasness, and noisiness. Transferability and generalizability are critical to address the OOD problem.

Several algorithms have been developed to solve multi-objective optimization problems in RL. Yang et al. introduced an envelope Q-learning method that can quickly adapt and solve new tasks with different preferences [98]. The capability of learning a single Q function over all preferences is important for personalized drug discovery. In another study, Chen et al. proposed a two-stage model [99] for multi-objective deep RL. At the first stage, a multi-policy soft actor-critic algorithm is applied to collaboratively learn multiple policies with different targets, in which each policy targets on a specific scalarized objective. At the second stage, a multi-objective covariance matrix adaptation evolution strategy is applied to fine-tune the policy-independent parameters.

*New methods to improve reward functions for OOD data.* A plethora of machine learning approaches including self-supervised learning, transfer learning, semi-supervised learning, meta-learning and their combinations have been recently developed to address out-of-distribution problems in compound screening in terms of chemicals, proteins and cell lines. Self-supervised learning has enjoyed a great success in Natural Language Processing, image recognition, and protein sequences modeling. Cai et al. have proposed a DIstilled Sequence Alignment Embedding (DISAE) transformer for predicting ligand binding to orphan receptors [100]. DISAE has been further extended to out-of-distribution prediction of receptor activities of ligand binding, specifically, agonist vs antagonist [101]. An out-of-cluster meta-learning algorithm has been proposed to explore dark chemical genomics space that includes all Pfam families [102]. Self-supervised learning and semi-supervised learning have also been applied to explore chemical space [103]. Transfer learning is particularly useful in predicting drug responses (both cell viability and gene expressions) for novel cell lines [104] or translating *in vitro* compound screens to clinical outcomes in patients [105]. These methods, when applied to reward functions, could improve the performance of RL in systems pharmacology.

Biophysics-based methods such as molecular dynamics (MD) simulation, quantum chemistry, and protein-ligand docking (PLD), can be directly applied to evaluate the chemical properties, and thus can be used as reward functions. MD simulation and quantum chemistry calculation are computationally expensive. Emergence of quantum computing may make it feasible to incorporate them directly into RL as reward functions [106]. With the advent of high-accuracy protein structural models, such as AlphaFold2 [15], it now becomes feasible to use PLD to predict ligand-binding sites and poses on dark proteins, on a genome-wide scale. However, PLD suffers from a high false-positive rate due to poor modeling of protein dynamics, solvation effects, crystallized waters, and other challenges [107]; often, small-molecule ligands will indiscriminately 'stick' to concave, pocket-like patches on protein surfaces. For these reasons, although AlphaFold2 can accurately predict many protein structures, the relatively low reliability of PLD still poses a significant limitation, even with a limitless supply of predicted structures [108]. Thus, the direct application of PLD remains a challenge for predicting ligand binding to dark proteins. Recently, Cai et al. have proposed an end-to-end sequence-structure-function learning framework PortalCG [17]: protein structure information is not used as a fixed input, but rather as an intermediate layer that can be tuned using various structural and functional information. PortalCG significantly outperforms the direct use of PLD for predicting ligand binding to dark proteins. Thus, it could be an effective strategy to incorporate biophysics domain-knowledge into deep learning frameworks as constraints or regularizations.

## 5. Conclusion

In spite of the great success of RL in GO board game, computer games, and other settings with well-defined environments, its application to drug discovery is still in its infancy. Current efforts in RL mainly focus on target-based drug design. Although RL is a promising technique, its actual value in the target-based drug discovery is overestimated in the current stage. The major hurdle comes from the reward function that is based on predicted binding affinities and needs to be obtained from either a machine learning approach or a physics-based scoring. Both of them are not reliable and accurate enough when applied to chemical compounds with novel structures. For a machine learning approach, it remains a great challenge to predict an OOD sample, i.e., a generated molecule that falls outside the distribution of chemicals in the training data for activity predictions. As a result, the structure of active compounds inferred from RL may not be significantly different from those in the training data. In terms of physical-based scoring, there is a lack of computationally efficient and accurate methods to model multiple factors including conformational dynamics, solvent effects, hydrogen bonding, entropies, crystallized waters, etc., which contribute to the protein-ligand binding affinity. Consequently, the scoring function is still suboptimal and unreliable. In the authors' humble opinion, existing RL approaches to de novo drug design are scarcely fruitful in regard to the structural novelty of chemical compounds without the significant improvement of efficiency and accuracy of binding affinity predictions.

Although current efforts in RL mainly center around the target-based drug design, systems pharmacology and personalized medicine have emerged as new paradigms in drug discovery. They have advantages over the conventional "one-drug-one-gene" approach when tackling multi-genic, heterogeneous diseases. On the other hand, systems pharmacology-oriented and personalized drug design imposes new challenges on RL-based drug design. Conceptually, systems pharmacology has not been fully appreciated by the pharmaceutical and biotechnology industry. Technically, there are few high-quality labeled data available for training a generalizable machine learning model to predict molecular phenotypic readouts suitable for systems pharmacology-oriented and personalized drug design. Thus, the OOD problem in systems pharmacology is more serious than the target-based drug design. Besides the OOD issue that incapacitates the reward function, the success of RL in systems pharmacology-oriented and personalized drug design needs to overcome additional roadblocks. Firstly, the generalizability and transferability are the central challenges for the deployment of RL in systems pharmacology as the environments are partially observable or changed dramatically (e.g., from cell lines to human tissues). Secondly, RL needs to optimize multiple dynamic and often conflict rewards in a non-stationary environment while the dynamic multi-objective optimization still remains as an unsolved research problem. Finally, it is necessary to integrate multiple heterogeneous, noisy, and high-dimensional omics data for successful systems pharmacology modeling, which is a challenging task under intensive investigations. Thus, the realization of the full potential of RL in drug discovery relies on not only new developments in RL but also advances in other fields that are beyond RL. Specifically, RL needs to be synergistically integrated with unsupervised/supervised machine learning, biophysics, systems biology, multi-omics technology, and quantum computing.